\documentclass[aps,prb,twocolumn, superscriptaddress]{revtex4-2}

\usepackage{graphicx}

\usepackage{multirow}

\usepackage{color} % use colored text in latex

\begin{document}

\title{Pressure induced enhancement of polar distortions in a metal, and implications on the Rashba spin-splitting}

\author{Evie Ladbrook} 
\affiliation{Department of Chemistry, University of Warwick, Gibbet Hill, Coventry, CV4 7AL, United Kingdom}

\author{Urmimala Dey} 
\affiliation{Centre for Materials Physics, Durham University, South Road, Durham DH1 3LE, United Kingdom}
\affiliation{Luxembourg Institute of Science and Technology (LIST), Avenue des Hauts-Fourneaux 5, L-4362, Esch-sur-Alzette, Luxembourg}

\author{Nicholas C. Bristowe} 
\affiliation{Centre for Materials Physics, Durham University, South Road, Durham DH1 3LE, United Kingdom}

\author{Robin S. Perry} 
\affiliation{London Centre for Nanotechnology and Department of Physics and Astronomy, University College London, London WC1E 6BT, United Kingdom}

\author{Dominik Daisenberger} 
\affiliation{Diamond Light Source Ltd, Harwell Science and Innovation Campus, Didcot OX11 0DE, United Kingdom}

\author{Mark R. Warren} 
\affiliation{Diamond Light Source Ltd, Harwell Science and Innovation Campus, Didcot OX11 0DE, United Kingdom}

\author{Mark S. Senn} 
\email{m.senn@warwick.ac.uk}
\affiliation{Department of Chemistry, University of Warwick, Gibbet Hill, Coventry, CV4 7AL, United Kingdom}

\date{\today}%

\begin{abstract}

Polar metals are an intriguing class of materials in which electric polarisation and metallicity can coexist within a single phase. The unique properties of polar metals challenge expectations, making way for the exploration of exotic phenomena such as unconventional magnetism, hyperferroelectric multiferroicity and developing multifunctional devices that can leverage both the materials electric polarization and its asymmetry in the spin conductivity, that arises due to the Rashba effect. Here, via a high pressure single crystal diffraction study, we report the pressure-induced enhancement of polar distortions in such a metal, Ca$_3$Ru$_2$O$_7$. Our DFT calculations highlight that naive assumptions about the linear dependency between polar distortion amplitudes and the magnitude of the Rashba spin splitting may not be generally valid.

\end{abstract}

\maketitle

Polar metals are an intriguing class of materials in which electric polarisation and metallicity can coexist within a single phase. The concept of a ferroelectric-like transition in a metal was first predicted in 1965 \cite{anderson_symmetry_1965} but not experimentally realised until such a transition was identified in metallic LiOsO$_{3}$ \cite{shi_ferroelectric-like_2013}. Several polar metals have been identified since then \cite{kim_polar_2016, puggioni_designing_2014, kolodiazhnyi_insulator-metal_2008}, with ferroelectric switching first being demonstrated in WTe$_{2}$ \cite{fei_ferroelectric_2018, sharma_room-temperature_2019}, defying assumptions that ferroelectricity and metallicity are mutually exclusive. The unique properties of polar metals do not only challenge expectations, making way for the exploration of exotic phenomena such as unconventional magnetism \cite{lei_comprehensive_2019}, hyperferroelectric multiferroicity \cite{luo_two-dimensional_2017} and unique topologies \cite{xu_discovery_2015}, but also open up the possibility for developing multifunctional devices \cite{puggioni_design_2015, puggioni_designing_2014, yao_manipulation_2017, liu_vertical_2019} that can leverage both electric polarisation, or ferroelectricity, and metallic conductivity. 

The absence of an inversion centre in combination with the spin orbit interaction can facilitate the Rashba interaction. As polar metals can satisfy both requirements, they provide an obvious platform to explore this effect. Although the Rashba interaction has most commonly been studied at surfaces and interfaces \cite{shanavas_electric_2014, bhowal_electric_2019}, where inversion symmetry is necessarily broken, the effect has also been demonstrated in bulk systems \cite{di_sante_electric_2013, ishizaka_giant_2011, yamauchi_bulk_2019}. This has implications to the field of spintronics \cite{manchon_new_2015}, where electron spins, rather than just the charge, may be manipulated by an electric field. 

\begin{figure}[h]
\includegraphics[width=8.5cm] {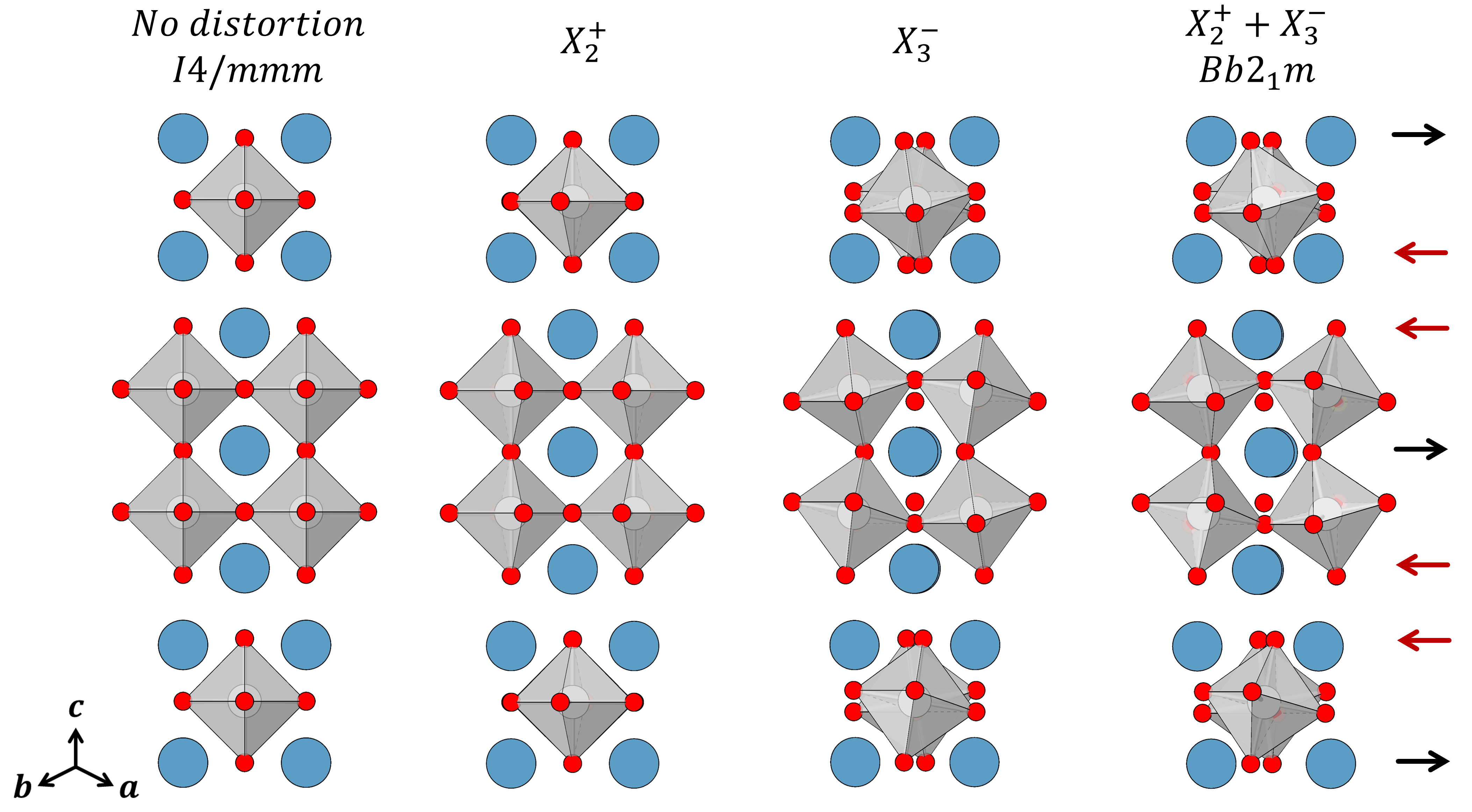}
\caption{\label{octahedra} $n$~=~2 Ruddlesden-Popper structure consisting of two perovskite layers seperated by a rock-salt-like layer with representation of the distortion of the theoretical high symmetry $I4/mmm$ to the orthorhombic $Bb2_{1}m$ phase through octahedral rotations (X$_2^+$) and tilts  (X$_3^-$). Arrows indicate the polar distortions, primarily consisting of Ca displacements.}
\end{figure}

In many polar metals, the polar mode arises through a ‘geometric mechanism’ as conventional chemical mechanisms for ferroelectricity are broadly incompatible with metallicity \cite{bhowal_polar_2023, kim_polar_2016, puggioni_designing_2014, benedek_ferroelectric_2016}. Typically, free electrons would screen the long-range interactions responsible for ferroelectricity, inhibiting formation of a permanent electric dipole. However, materials in which a polar mode arises due to a structural instability may circumvent this. One mechanism is via a ‘hybrid improper’ mechanism \cite{benedek_hybrid_2011}, prototypical to $n$~=~2 Ruddlesden-Popper (RP) phase Ca$_{3}$Mn$_{2}$O$_{7}$, where polarisation arises due to the trilinear coupling between two non-polar structural distortions, tilting and rotation of the MnO$_{6}$ octahedra in the case of Ca$_{3}$Mn$_{2}$O$_{7}$, with a polar distortion mode, as shown in Fig \ref{octahedra}. 

Ca$_{3}$Ru$_{2}$O$_{7}$ is a member of the $n$~=~2 RP family, crystallizing in the polar $Bb2_{1}m$ space group (non-standard setting of $Cmc2_{1}$), however, it has a metallic ground state \cite{cao_observation_1997, lei_observation_2018}. Although ferroelectric switching cannot be achieved by application of an external electric field, a recent study has revealed a polar domain structure consisting of 90- and 180-degrees domain walls with ferroelastic domains that can be switched by applied uniaxial strain \cite{lei_observation_2018}. Ca$_{3}$Ru$_{2}$O$_{7}$ also undergoes a spin reorientation transition where, upon cooling from 60 to 48 K, the spins reorient from aligning along the crystallographic $a$ axis, aligning ferromagnetically within each perovskite layer and antiferromagnetically between them, to be parallel to $b$ \cite{cao_observation_1997, bao_spin_2008, bohnenbuck_magnetic_2008}. The coupling between the spin ordering direction and the polar crystallographic axis via a strong spin-orbit interaction induces the Rashba-like spin splitting in momentum space \cite{markovic_electronically_2020}. Very recently, it has been shown that this transition can be induced by application of uniaxial strain which emphasizes how the structural, magnetic and electronic degrees of freedom are intricately coupled \cite{dashwood_strain_2023}. 

Similarly, pressure can be used to directly probe the interplay between the structural, octahedral rotations and tilts \cite{ramkumar_octahedral_2021}, and polar distortions with a view to understanding how the Rashba spin splitting might be tuned in polar metals. Although it is generally understood that the initial application of hydrostatic pressure suppresses polarisation in conventional proper ferroelectrics, such as BaTiO$_{3}$ \cite{ishidate_phase_1997, bousquet_first-principles_2006}, as they tend to undergo a phase transition to a non-polar state, this does not necessarily hold true for all ferroelectrics. For example, in the case of the hybrid improper mechanism, we have recently shown, by powder X-ray diffraction in combination with density functional theory calculations (DFT), that polarisation can be enhanced in the insulator Ca$_{3}$Ti$_{2}$O$_{7}$ by pressure \cite{clarke_pressure-dependent_2024}. 

The prospect of coupling spin transport properties to external stimuli like pressure via internal structural degrees of freedom, such as octahedral rotations, and their coupling to polar modes, is intriguing. However, naively, we might assume that the amplitude of the Rashba effect has a linear dependence on the polar mode, as shown for surfaces, interfaces and bulk systems \cite{leppert_electric_2016, da_silveira_rashba-dresselhaus_2016, hanakata_strain-induced_2018, noel_non-volatile_2020, jin_giant_2022}, however, there are indications in the literature that this situation may be somewhat more complex \cite{wang_anomalous_2022}. 

Here we report the result of our high-pressure single crystal diffraction studies on Ca$_{3}$Ru$_{2}$O$_{7}$ that show that the structural ingredients, necessary for enhancing the amplitude of the polar mode in a metal, are systematically enhanced up to 15 GPa. Our first-principles calculations within density functional theory (DFT) reveal that, contrary to expectations, the enhancement of the polar mode manifests itself in a decrease in spin splitting in momentum space. 

Single crystal samples were selected from the same growth batch as those which demonstrated incommensurate magnetism at 50K \cite{faure_magnetic_2023}. All samples were confirmed as single phase and good quality via magnetisation, resistivity and SEM/EDX measurements.

A single crystal of dimensions 50~$\times$~30~$\times$~15 $\mu$m was loaded into a LeToullec diamond anvil cell (DAC) equipped with Boehler-Almax anvils with 400 $\mu$m culets and a rhenium gasket which was pre-indented to 50 $\mu$m with a 250 $\mu$m diameter sample chamber eroded into the indent. Helium was used as a pressure transmitting medium to ensure the sample was compressed under hydrostatic pressure. A ruby sphere was used as pressure indicator \cite{ruby}. Single crystal X-ray diffraction measurements were performed using a four-axis Newport diffractometer equipped with a Dectris Eiger CdTe detector, operating at a wavelength of $\lambda$ = 0.4859 \r{A} (corresponding to a beam energy of approximately 25.5 keV), in Experimental Hutch 2 (EH2) at Beamline I19 of the Diamond Light Source. 

Indexing, integration and refinalisation were performed using CrysAlisPRO with a spherical absorption correction. The structure was initially solved for the ambient pressure collection using SHELXT \cite{sheldrick_shelxt_2015} and subsequent pressures by isomorphous replacement, and then refined to convergence using SHELXL \cite{shelxl} implemented through Olex2 \cite{dolomanov_olex2_2009}. The presence of inversion twinning was accounted for via the ([-1 0 0], [0 -1 0], [0 0 1]) twin law for which the batch scale factor (BASF) was fixed to 0.50. 

High pressure single crystal X-ray diffraction data were collected between 0 and 14.5(2) GPa. Fig 2 shows how the experimentally refined $a$ and $b$ lattice parameters and volume evolve with pressure compared to those calculated from DFT. Full computational details can be found in the Supplemental Material (SM) \cite{Supplemental_Material}. In both cases, the lattice parameters and volume decrease smoothly with pressure. Experimentally, we find a slightly greater compressibility in \textit{b}, and smaller in \textit{a}, compared to that calculated by DFT.  However, this could be because the DFT calculations were performed at 0 K whilst experimental measurements were carried out at room temperature. Compressibility parameters and Birch-Murnaghan coefficients were calculated using PASCal \cite{cliffe_pascal_2012} and are tabulated in the Supplemental Material [Tables S2 and S3 \cite{Supplemental_Material}]. 

\begin{figure}[h]
\includegraphics[width=8.5cm] {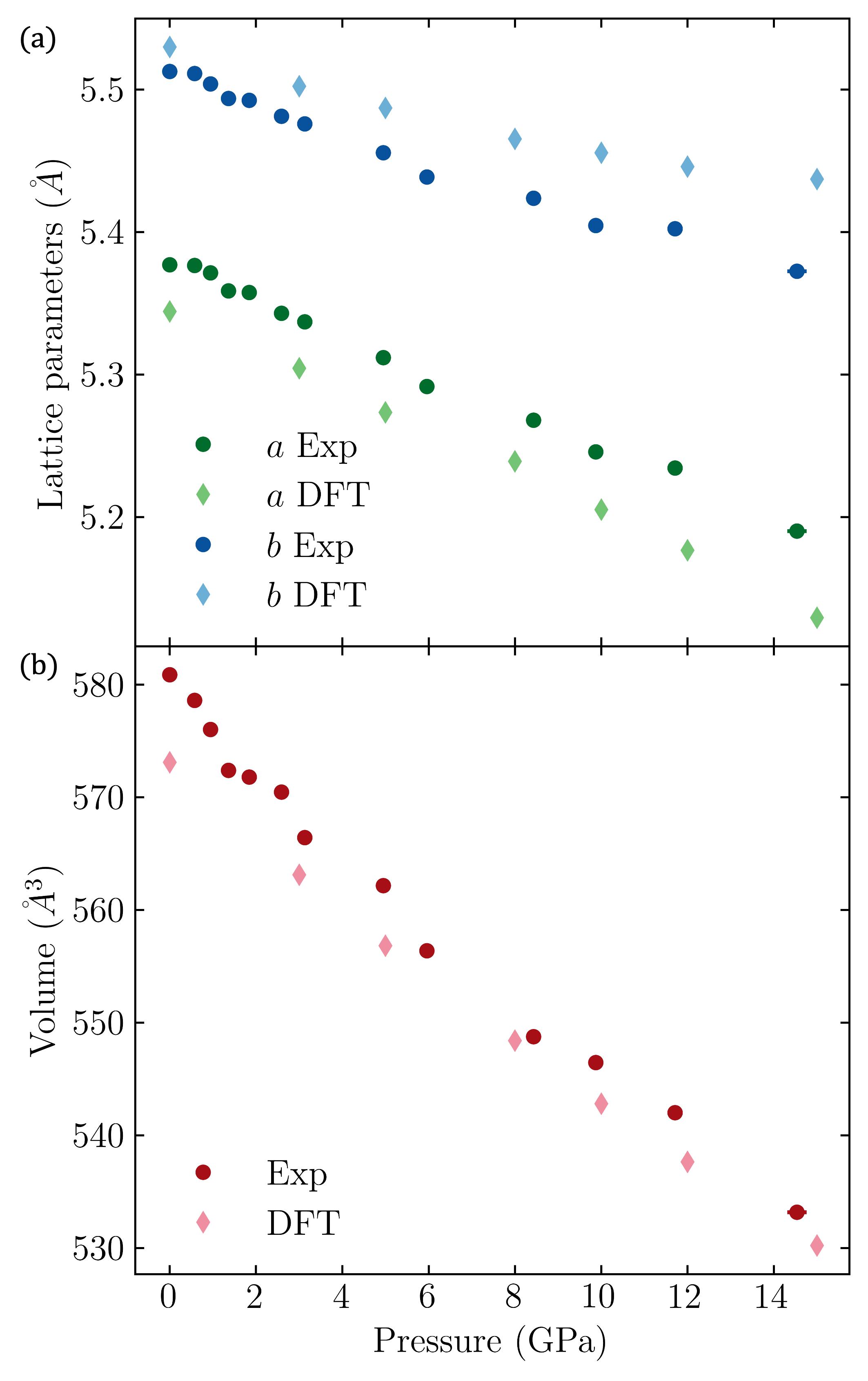}
\caption{\label{volume} (a) $a$ and $b$ lattice parameters and (b) unit cell volume extracted from variable pressure single crystal diffraction data and DFT calculations. }
\end{figure}

To explore the evolution of the atomic level structure, the experimental and DFT-relaxed structures were then decomposed in terms of symmetry-adapted displacements using ISODISTORT \cite{campbell_b_j_isodisplace_2006, isotropy}. The $Bb2_{1}m$ phase is related to the theoretical $I4/mmm$ aristotype by an in-phase rotation and an out-of-phase tilt of the TiO$_{6}$ octahedra which transform as irreducible representations X$_2^+$ and X$_3^-$, respectively. These couple to the polar displacement, which transforms as irreducible representation $\Gamma$$_5^-$, via the trilinear coupling mechanism. The resulting pressure dependent symmetry distortion mode amplitudes are shown in Fig 3. 

\begin{figure}[h]
\includegraphics[width=8.5cm] {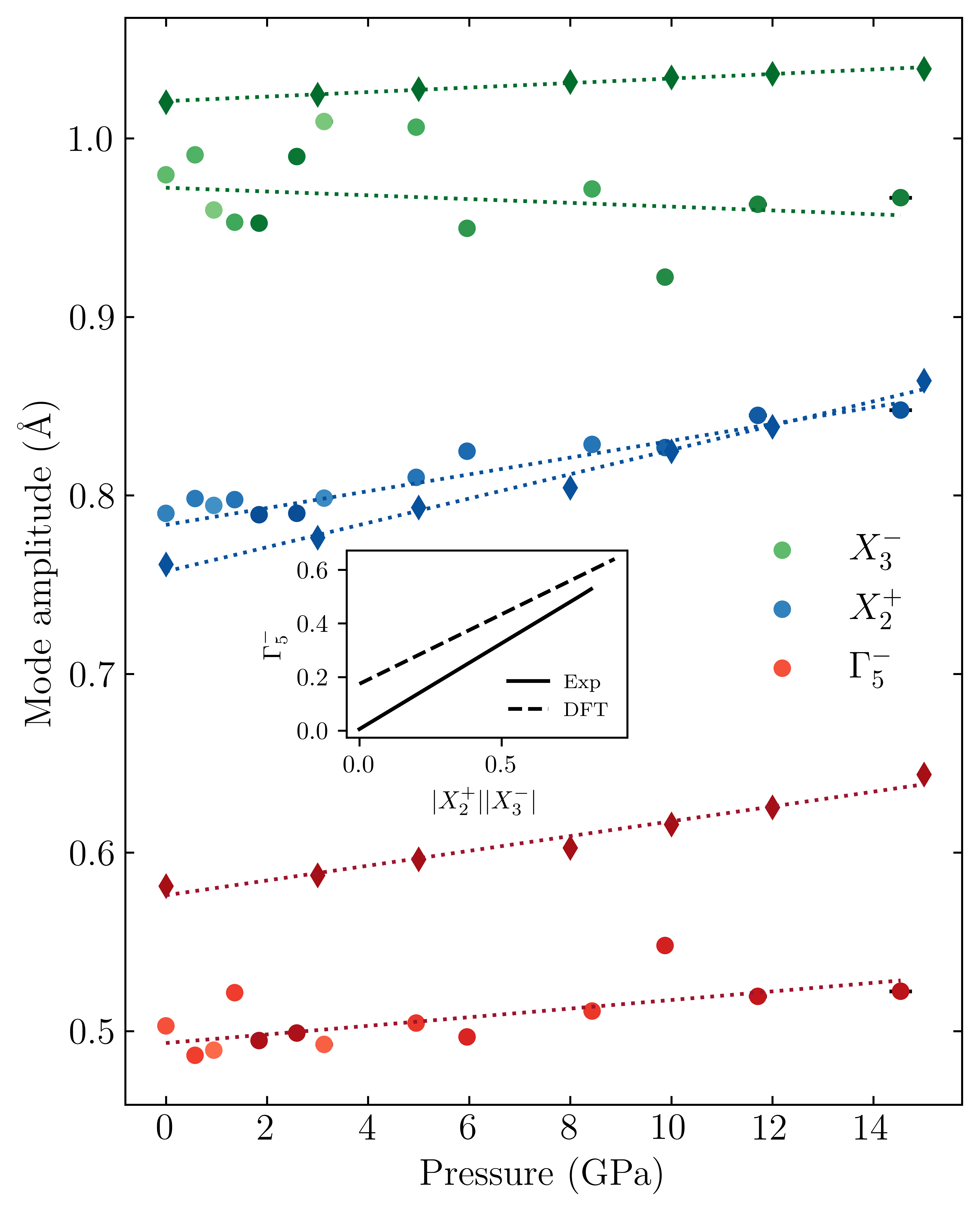}
\caption{\label{modes} Distortion mode amplitudes as a function of pressure for experimental (circles) and DFT calculated (diamonds) results. For the experimental results, a weighted linear regression was performed based on the R$_1$ statistic of the refined structures, with darker colours indicating a lower R$_1$ value. The inset shows a plot of the product of the amplitude of the rotation (X$_2^+$) and tilt (X$_3^-$) modes, from these fits, against the polar mode ($\Gamma$$_5^-$) amplitude.}
\end{figure}

The experimentally determined mode amplitudes show an overall steady increase in the amplitude of the rotation, X$_2^+$, whilst the amplitude of the tilt, X$_3^-$, remains relatively constant as the pressure increases. Despite some fluctuations in the experimentally determined mode amplitudes due to restrictions concerning the experimental set up, such as narrow opening angle of the DACs, these results show a good level of agreement with the DFT calculated mode amplitudes. In accordance with the trilinear coupling mechanism, the amplitude of the polar distortion, $\Gamma$$_5^-$, should be linearly proportional to the product of the amplitudes of the driving order parameters, X$_2^+$ and X$_3^-$, which is demonstrated in the inset of Fig \ref{modes}, showing a linear trend with an intercept at zero \cite{benedek_hybrid_2011}. This confirms that, for the experimental data, the polarisation is driven solely by the hybrid improper mechanism. However, for the DFT calculated results, the $\Gamma$$_5^-$ amplitude is in excess of that expected by the trilinear mechanism. This suggests that there is a small proper contribution to the polarisation, which, as shown in Fig. S3 \cite{Supplemental_Material}, vanishes as the pressure increases above 10 GPa. This could be for a variety of reasons such as the difference in volume of the DFT unit cell, dependence on the pseudopotentials or exchange correlation functional or it could be a real discrepancy between 0 K DFT and our room temperature diffraction experiments. However, in general, the level of agreement, especially considering both the complexity of the electronic structure of these materials and the challenges of performing high resolution x-ray diffraction at these pressures, is very good. The key point from both the experimental and DFT work is that that octahedral rotations are enhanced at a greater rate than the octahedral tilts are suppressed, resulting in an increase in the polarisation via the trilinear mechanism. While this pressure induced enhancement would be unexpected for a proper ferroeletric, we have recently predicted, via DFT calculations, a similar effect in insulating hybrid improper ferroelectric Ca$_3$Ti$_2$O$_7$ \cite{clarke_pressure-dependent_2024}, and thus our experimental and DFT results are suggestive that this phenomena may be much more wide spread.

Motivated by our ability to enhance the polar distortion in this material as a function of pressure, we now turn our attention to the effect such enhancements might have on the band structure of these materials.

\begin{figure}[h]
\includegraphics[scale=1.32]{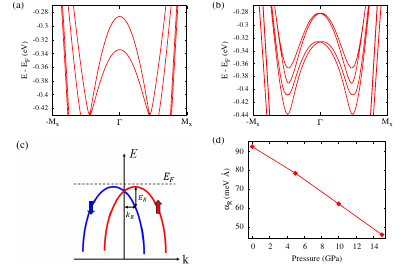}
\caption{\label{Rashba} Electronic band structure calculated at ambient pressure (a) without and (b) with SOC. $E_F$ is the Fermi energy. (c) Schematic diagram illustrating the definition of Rashba parameter, where $k_R$ is the distance between the crossing point of the Rashba spin-split bands and VBM, and $E_R$ is the corresponding energy difference. Red and blue arrows indicate the up and down spin-character of the bands, respectively. (d) Evolution of the Rashba parameter $\alpha_R$ in Ca$_3$Ru$_2$O$_7$ as a function of pressure.}
\end{figure}

Figs.~\ref{Rashba}(a-b) present the band structures calculated for the ambient-pressure $Bb2_1m$ structure of Ca$_3$Ru$_2$O$_7$ along the -$M_x-\Gamma-M_x$ direction with and without spin-orbit coupling (SOC). We observe a sizeable Rashba splitting of the bands around the $\Gamma$-point near the valence band maximum (VBM) when SOC is turned on. 

To quantify the magnitude of the momentum-dependent Rashba splitting, we compute the Rashba parameter defined as $\alpha_R = 2E_R/k_R$, where $k_R$ is the distance between the crossing point of the Rashba-split bands and VBM, and $E_R$ is the corresponding energy splitting~\cite{leppert_electric_2016, da_silveira_rashba-dresselhaus_2016}, as shown in Fig.~\ref{Rashba}(c). Surprisingly, the Rashba parameter calculated for the fully relaxed structures at different pressures reveal a monotonic decrease in $\alpha_R$ with increasing pressure [Fig.~\ref{Rashba}(d)]. Since the polar mode amplitude in Ca$_3$Ru$_2$O$_7$ increases with pressure [Fig \ref{modes}], this observation contradicts the conventional expectation that the Rashba parameter should scale with the polar mode amplitude \cite{leppert_electric_2016, da_silveira_rashba-dresselhaus_2016, hanakata_strain-induced_2018, noel_non-volatile_2020, jin_giant_2022}. In fact, the increase in the amplitude of the polar mode appears to be the primary cause of the reduction in the Rashba parameter. As detailed in S7 \cite{Supplemental_Material}, when the polar mode is absent, leaving only rotation and tilt modes, which by themselves break inversion symmetry, the Rashba parameter is enhanced compared to its value when the polar mode is present. Unpicking the multiple complexities of the pressure-dependent structure, and band structure, represent substantial further work, but our result is strongly suggestive that the magnitude of the structural polarisation is an insufficient proxy for the k-space spin-polarisation dependence.

In conclusion, we have performed a detailed high pressure investigation into the pressure dependence of the hybrid improper ordering parameters in polar metal, Ca$_3$Ru$_2$O$_7$.  Our results show that, contrary to the situation predominantly observed for proper ferroelectrics, hydrostatic pressure actually acts to enhance the polar distortions in this materials. This comes about due to the increase in the magnitude of RuO$_{6}$ octahedral rotations, which form one of the primary order parameters of the hybrid improper mechanism. Curiously, despite the enhanced magnitude of the polar mode at high pressures, our theoretical calculations show that, counterintuitively, this would lead to a decrease in the Rashba spin-splitting. Thus, our results highlight the Rashba effect in polar metals as a distinctive phenomenon worthy of further investigation.

\begin{acknowledgments}
E.L. thanks the University of Warwick for a PhD studentship through the Warwick Centre for Doctoral Training in Analytical Science and thanks Matt Edwards for assistance in the collection of high pressure X-ray diffraction data.
U.D. and N.C.B. acknowledge the Leverhulme Trust for a research project grant (Grant No. RPG-2020-206). This work used the ARCHER2 UK National Supercomputing Service (https://www.archer2.ac.uk) \cite{archer} and the Hamilton HPC Service of Durham University.
M.S.S. acknowledges the Royal Society for a fellowship (UF160265 $\&$ URF$\backslash$R$\backslash$231012) and EPSRC grant “Novel Multiferroic Perovskites through Systematic Design” (EP/S027106/1) for funding. The initial high pressure diffraction studies were performed via the the
Warwick X-ray Research Technology Platform. We acknowledge Diamond Light Source for time on beamline I19 under proposal CY36775.
\end{acknowledgments}

\bibliography{apssamp}

\clearpage
\newpage

\onecolumngrid

\section*{SUPPLEMENTAL MATERIAL}
\setcounter{page}{1}
\setcounter{figure}{0}
\setcounter{table}{0}
\setcounter{section}{0}
\renewcommand{\thepage}{S\arabic{page}}
\renewcommand{\thesection}{S\arabic{section}}
\renewcommand{\thetable}{S\arabic{table}}
\renewcommand{\thefigure}{S\arabic{figure}}
\newcounter{SIfig}
\renewcommand{\theSIfig}{S\arabic{SIfig}}

\begin{table}[ht!]
  \setlength{\tabcolsep}{6.0pt}
  \caption{Lattice parameters and refinement statistics }
  \label{tab1}
  \centering
  \begin{tabular}{|c|c|c|c|c|c|c|c|c|c|c|}
    \hline  \hline \rule{0pt}{1.2\normalbaselineskip}
Pressure (GPa) & $a$ (Å)      & $b$ (Å)      & $c$ (Å)     & R$_{1}$   & wR$_{2}$   & R$_{int}$ & GooF   \\ \hline 
0.00(5)        & 5.3770(5)  & 5.5126(13) & 19.60(4)  & 5.51 & 11.76 & 2.70  & 1.23   \\ \hline 
0.58(5)        & 5.3764(5)  & 5.5112(12) & 19.53(6)  & 5.00 & 12.89 & 2.88 & 1.203  \\ \hline 
0.95(9)  & 5.3713(5)  & 5.5038(15) & 19.48(6)  & 6.58 & 15.28 & 2.48 & 1.222   \\ \hline 
1.36(9)  & 5.3587(10) & 5.494(2)   & 19.44(10) & 4.61 & 12.90 & 3.02 & 1.248   \\ \hline 
1.84(13)     & 5.3576(5)  & 5.4924(10) & 19.43(7)  & 1.98 & 5.54  & 1.91 & 1.272   \\ \hline 
2.59(13)     & 5.3429(7)  & 5.4812(17) & 19.48(8)  & 1.98 & 6.32  & 1.85 & 1.230    \\ \hline 
3.13(17)     & 5.3370(7)  & 5.476(2)   & 19.38(8)  & 6.13 & 13.12 & 4.17 & 1.344   \\ \hline 
4.95(13)    & 5.3119(6)  & 5.4556(14) & 19.40(6)  & 4.54 & 10.14 & 3.18 & 1.151   \\ \hline 
5.96(13)   & 5.2915(8)  & 5.439(2)   & 19.33(9)  & 3.79 & 9.45  & 2.17 & 1.115  \\ \hline 
8.43(13)    & 5.2680(9)  & 5.424(3)   & 19.21(11) & 4.50  & 10.37 & 1.56 & 1.220    \\ \hline 
9.88(14)     & 5.2457(6)  & 5.405(2)   & 19.27(7)  & 3.26 & 8.49  & 2.19 & 1.220    \\ \hline 
11.71(18)      & 5.2344(6)  & 5.402(2)   & 19.17(7)  & 2.75 & 7.53  & 1.80  & 1.268 \\\hline 
14.5(2)      & 5.1900(5)  & 5.3724(18) & 19.12(7)  & 2.43 & 6.30   & 1.11 & 1.192   \\
\hline 
\hline 
\end{tabular}
\end{table}

\begin{table}[ht!]
  \setlength{\tabcolsep}{6.0pt}
  \caption{Compressibility parameters calculated using PASCal for experimental and DFT calculated results. }
  \label{tab3}
  \centering
  \begin{tabular}{|c|c|c|c|c|c|c|c|}
    \hline  \hline\rule{0pt}{1.2\normalbaselineskip}
    \multirow{2}{2em}{}&\multirow{2}{2em}{Axes}&\multirow{2}{4em}{K (TPa\textsuperscript{-1})}&\multirow{2}{4em}{$\sigma$K (TPa\textsuperscript{-1})}& \multicolumn{4}{c|} {Empirical parameters} \\ \cline{5-8}
    &&& & $\varepsilon$ & $\lambda$ & P\textsubscript{c} & v \\ \hline 
    \multirow{4}{2em}{Exp} & $a$ & 2.6997 & 0.1342 & 0.0008 & -0.0037 & 0.4024 & 0.8482 \\ \cline{2-8}
    & $b$ & 2.1150 & 0.1046 & 0.0006 & -0.0038 & -0.4897 & 0.7306 \\ \cline{2-8}
    & $c$ & 1.6427 & 0.2274 & 0.0 & -0.0037 & 0.5765 & 0.6314 \\ \cline{2-8}
    & V & 6.0733 & 0.1915 & \multicolumn{4}{c|}{} \\ \hline
    \multirow{4}{2em}{DFT} & $a$ & 2.7334 & 0.0045 & 0.1231 & 0 & -497.648 & 9.6142 \\ \cline{2-8}
    & $b$ & 1.0187 & 0.0494 & 17.2944 & -17.2633 & -7.3077 & 0.0009 \\ \cline{2-8}
    & $c$ & 1.2621 & 0.0403 & 8.567 & -8.476 & -17.4137 & 0.0037 \\ \cline{2-8}
    \rule{0pt}{1.0\normalbaselineskip} & V & 4.9735 & 0.1321 & \multicolumn{4}{c|}{} \\ 
    \hline 
    \hline
\end{tabular}
\end{table}

\begin{table}[ht!]
  \setlength{\tabcolsep}{6.0pt}
  \caption{Second-order Birch-Murnaghan coefficients calculated using PASCal. }
  \label{tab4}
  \centering
  \begin{tabular}{|c|c|c|c|c|c|c|c|}
    \hline  \hline \rule{0pt}{1.2\normalbaselineskip}

    {} & B\textsubscript{0} (GPa) & $\sigma$B\textsubscript{0} (GPa) & V\textsubscript{0} (\AA\textsuperscript{3}) & $\sigma$V\textsubscript{0} (\AA\textsuperscript{3}) & B' & $\sigma$B' & P\textsubscript{c} (GPa) \\ \hline

    Exp & 135.3161 & 4.8459 & 581.0391 & 0.0934 & 4 & N/A & 0 \\ \hline
    DFT & 165.2915 & 0.4932 & 573.0904 & 0.0787 & 4 & N/A & 0 \\ 

    \hline 
    \hline
\end{tabular}
\end{table}

\newpage

\section{Computational Details}
We performed first-principles simulations based on density functional theory (DFT) implemented within the VASP code~\cite{VASP1,VASP2}, version 6.3.2. PBEsol variant of generalised gradient approximation (GGA)~\cite{PBEsol} was employed as the exchange-correlation functional to accurately describe the structural properties of bulk Ca$_3$Ru$_3$O$_7$. We used PAW pseudopotentials (PBE, version 5.4)~\cite{pseudo1,pseudo2} with 10 valence electrons for Ca ($3s^23p^24s^2$), 14 for Ru ($4p^64d^75s^1$), and 6 for O ($2s^22p^4$). An effective on-site Hubbard parameter $U_{\text{eff}}$ = 1.2 eV was utilised to account for the correlation effects of the $4d$ electrons in Ru~\cite{HubbardU},  reproducing the ambient-pressure structural and magnetic properties, as well as the band dispersion, in close agreement with previous experimental observations. Spin-orbit coupling (SOC) was included self-consistently within VASP. A plane wave cutoff of 800 eV and $k$-mesh grid of $8 \times 7 \times 2$ were used to resolve the total energies, forces, and stresses within 1 meV/f.u., 1 meV/{\AA}, and 0.1 GPa, respectively. An energy convergence criterion of 10$^{-9}$ eV was used for all the calculations and full structural relaxations were carried out till the Hellmann–Feynman forces on each atom were less than 1 meV/{\AA}. Symmetry mode analyses were performed using the web-based ISOTROPY software suite~\cite{isotropy}. Crystal structures as well as magnetic configurations were visualised with VESTA~\cite{VESTA}.

\section{Determination of Magnetic Ground State from DFT}
\begin{figure}[ht!]
\includegraphics[scale=1.3]{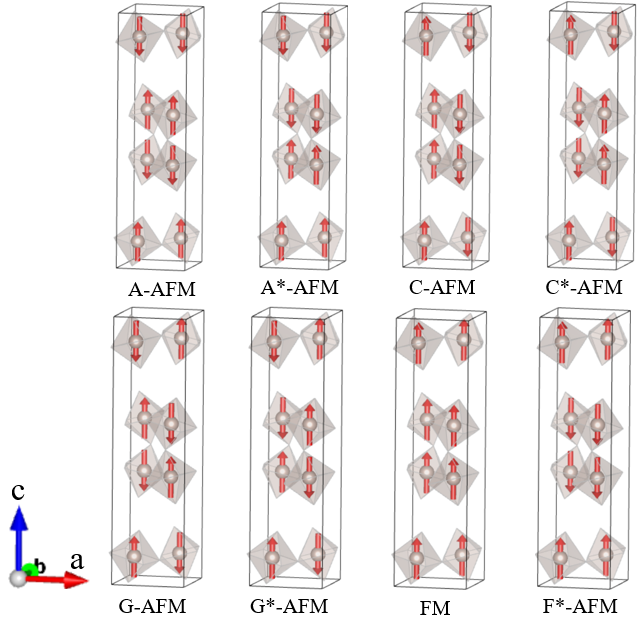}
    \caption{Collinear spin configurations considered for determining the ground state magnetic structure of Ca$_2$Ru$_2$O$_7$. The grey spheres represent the Ru atoms and the red arrows indicate the spin magnetic moments on the Ru$^{4+}$ ions.}
\refstepcounter{SIfig}\label{magnstruct}
\end{figure}

To investigate the ground state magnetic configuration, we consider a ferromagnetic (FM) and 7 different antiferromagnetic (AFM) collinear spin arrangements shown in Fig.~\ref{magnstruct}. Our collinear calculations without SOC reveal that F*-AFM corresponds to the ground state magnetic structure in agreement with previous theoretical and experimental findings~\cite{cao_observation_1997, bao_spin_2008, bohnenbuck_magnetic_2008,markovic_electronically_2020,Rondinelli2020,Leon2024}. The obtained magnetic moment of 1.41 $\mu_{\rm{B}}$ per Ru atom indicates the low-spin state of Ru$^{4+}$ ions~\cite{Rondinelli2020,Leon2024}. Furthermore, our $0$ K non-collinear calculations with SOC confirm the crystallographic $b$-axis as the easy direction for magnetization~\cite{cao_observation_1997, bao_spin_2008, bohnenbuck_magnetic_2008,markovic_electronically_2020,Rondinelli2020,Leon2024}. 

\begin{table}[ht!]
  \setlength{\tabcolsep}{6.0pt}
  \caption{Relative energies (meV/f.u.) of the different collinear spin configurations considered in our DFT+$U$ calculations. $U_{\rm{eff}}$ = 1.2 eV is used for the Ru-$5d$ orbitals without SOC.}
  \label{tab-coll}
  \centering
  \begin{tabular}{|c|c|c|c|c|c|c|c|}
    \hline  \hline\rule{0pt}{1.2\normalbaselineskip}
    A-AFM&A*-AFM& C-AFM& C*-AFM &G-AFM&G*-AFM& FM & F*-AFM \\ \hline 
    \rule{0pt}{1.0\normalbaselineskip}       98.46&99.53&196.75&198.51&190.12&191.64&1.13&0.00\\ 
    \hline 
    \hline
\end{tabular}
\end{table}

\begin{table}[ht!]
  \setlength{\tabcolsep}{6.0pt}
  \caption{Relative energies (meV/f.u.) of the F*-AFM magnetic configuration with different spin axes including SOC. }
  \label{tab-easydir}
  \centering
  \begin{tabular}{|c|c|c|}
    \hline
    \hline\rule{0pt}{1.2\normalbaselineskip}
    along $a$&along $b$&along $c$\\\hline 
    \rule{0pt}{1.0\normalbaselineskip} 
      2.38&0.00&6.62\\
    \hline 
    \hline
\end{tabular}
\end{table}

\newpage
\section{Band structure at ambient pressure}
\begin{figure}[ht!]
\includegraphics[scale=2.5]{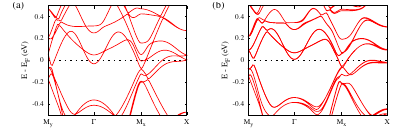}
    \caption{Electronic band structure of Ca$_3$Ru$_2$O$_7$ at ambient pressure along the $M_y(0, 0.5, 0) - \Gamma(0, 0, 0) - M_x(0.5, 0, 0)- X(0.5, 0.5, 0)$ path calculated (a) without and (b) with SOC using a $U_{\rm{eff}}$ value of 1.2 eV. }
\refstepcounter{SIfig}\label{BS_P0}
\end{figure}

Fig.~\ref{BS_P0} shows the electronic band structures computed for the $Bb2_1m$ structure of Ca$_3$Ru$_2$O$_7$ at ambient pressure with $U_{\rm{eff}}$ = 1.2 eV. We observe an overall metallic band structure with and without SOC. However, inclusion of SOC additionally opens up a local gap around the $\Gamma$-point observed in previous angle-resolved photoemission spectroscopy (ARPES) measurements~\cite{markovic_electronically_2020,Horio2021} highlighting the importance of SOC interaction in accurately describing the band structure at ambient pressure. 

Our band structure results contradict previous PBEsol+$U$ studies that reported a band crossing around $\Gamma$ with $U_{\rm{eff}} = 1.2$ eV~\cite{Rondinelli2020,Leon2024}. To verify our findings, we have performed a full structural relaxation of the $Bb2_1m$ structure with SOC, which yielded an identical band structure featuring a partial gap at the Fermi level around the $\Gamma$-point. Thus, our results demonstrate that PBEsol functional with $U_{\rm{eff}}$ = 1.2 eV reliably reproduces the ambient-pressure band dispersion of Ca$_3$Ru$_2$O$_7$ closer to the experimental observations. 

\newpage
\section{Evolution of the proper vs. improper component of the polar mode with pressure}
\begin{figure}[ht!]
\includegraphics[scale=0.4]{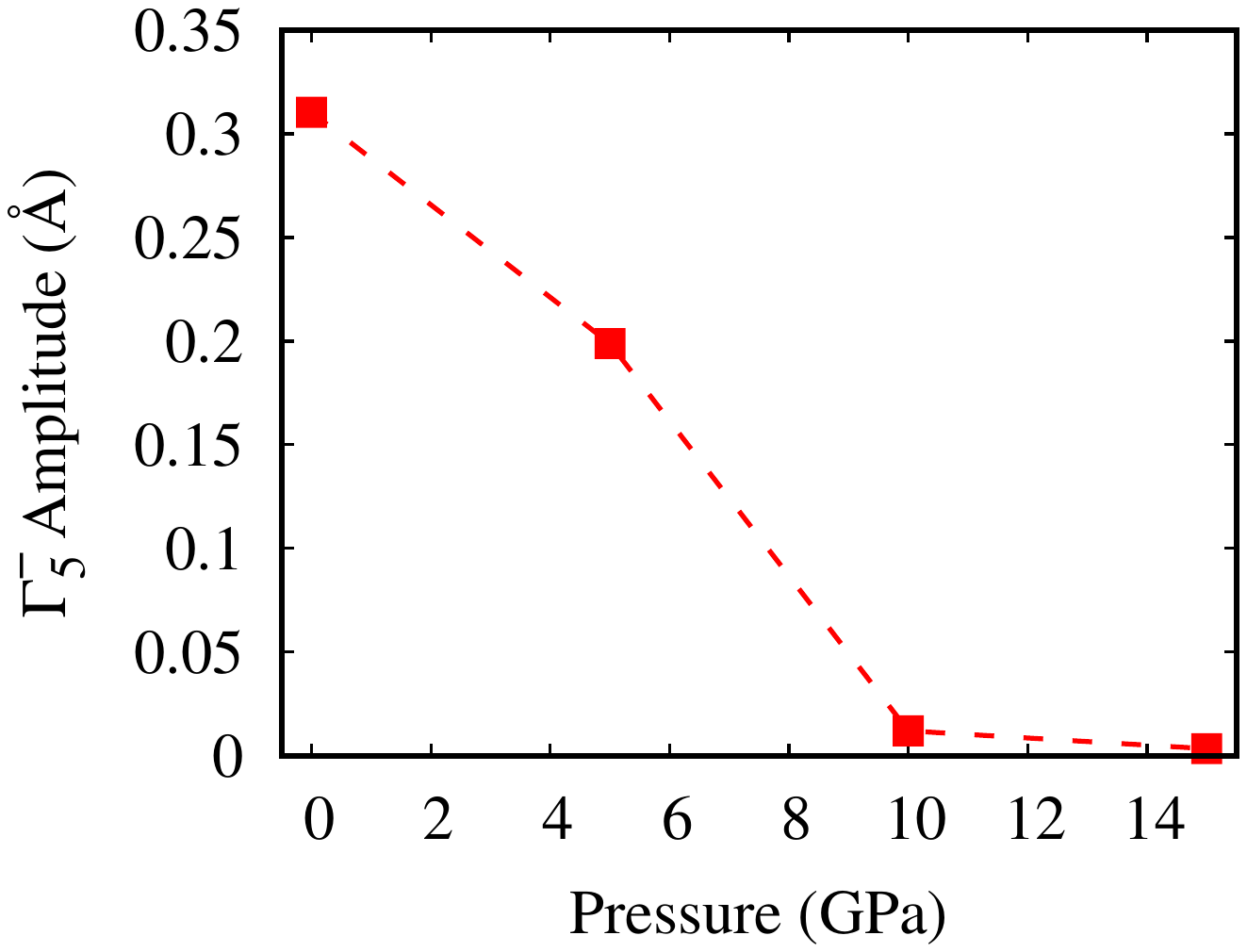}
    \caption{Variation of the polar mode amplitude with pressure in the $Fm2m$ phase.}
\refstepcounter{SIfig}\label{Fm2m}
\end{figure}

As discussed in the main text, at low pressure, amplitude of the polar mode transforming as the $\Gamma^-_5$ irrep in the DFT-calculated structures is in excess of that expected by the trilinear mechanism, suggesting that there is a small
proper contribution to the polarisation at low pressure. To confirm this small additional proper component to the polarisation, we compute the polar mode amplitude for an $Fm2m$ structure containing only the $\Gamma^-_5$ mode as a function of pressure. We show in Fig.~\ref{Fm2m} that the proper component vanishes when the hydrostatic pressure increases above 10 GPa.

\newpage

\section{Evolution of band structure with pressure}
\begin{figure}[ht!]
\includegraphics[scale=2.5]{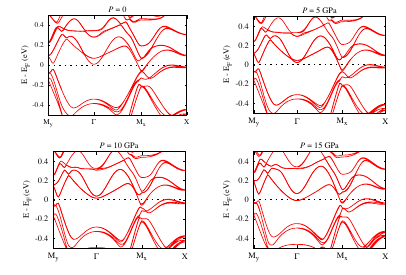}
    \caption{Electronic band structure of Ca$_3$Ru$_2$O$_7$ as a function of pressure. SOC is included in all the calculations. }
\refstepcounter{SIfig}\label{BS}
\end{figure}

\section{Evolution of the magnetic moment with pressure}
\begin{figure}[ht!]
\includegraphics[scale=0.6]{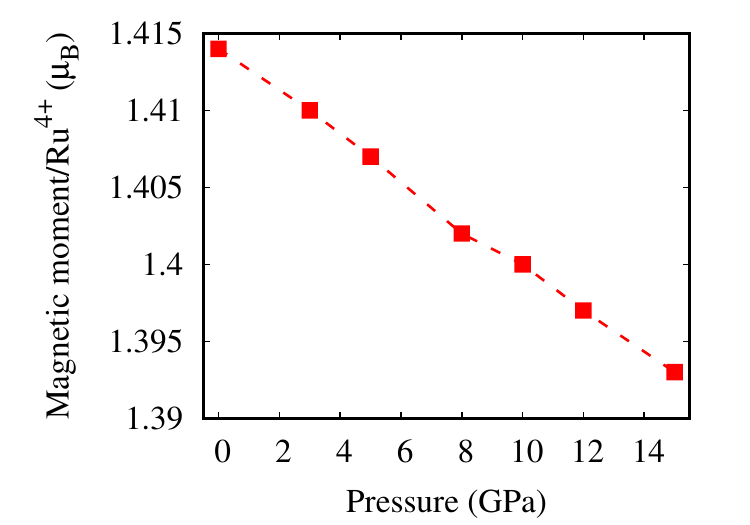}
    \caption{Spin magnetic moment on the Ru$^{4+}$ ions as a function of pressure obtained from collinear calculations without SOC.}
\refstepcounter{SIfig}\label{moment}
\end{figure}
Our non-collinear calculations including SOC confirm that the $b$-axis remains the easy magnetisation direction throughout the $0-15$ GPa pressure range considered in our work.

\section{Change in the Rashba spin-splitting with pressure}
\begin{figure}[ht!]
\includegraphics[scale=2.5]{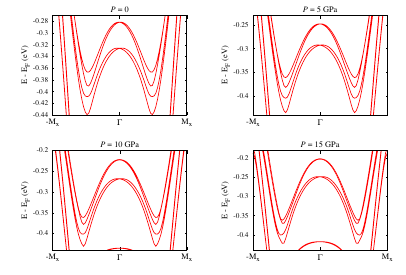}
    \caption{Electronic band structures at different hydrostatic pressure, showing the decrease in the Rashba spin-splitting at the VBM as a function of pressure. }
\refstepcounter{SIfig}\label{BS_Rashba}
\end{figure}

\begin{figure}[ht!]
\includegraphics[scale=0.97]{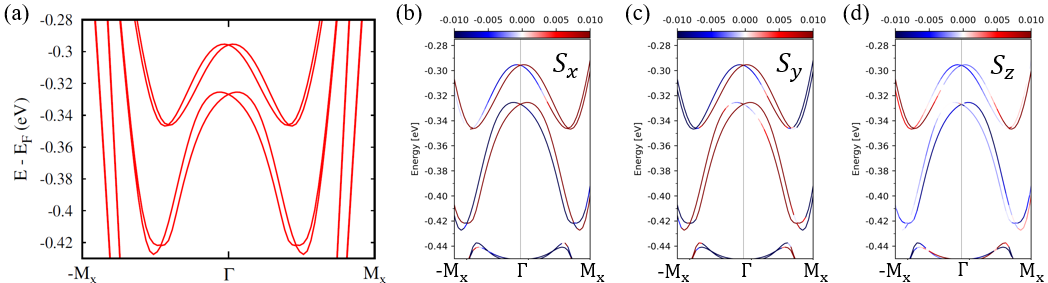}
    \caption{(a) Band structure at ambient pressure calculated by setting the polar mode amplitude to zero (tilts only). (b-d) Corresponding spin-resolved band structures confirming the pure Rashba character of the bands at the VBM.}
\refstepcounter{SIfig}\label{BS_mode}
\end{figure}

\newpage
To understand the decrease in the Rashba splitting with pressure, depicted in Fig.~\ref{BS_Rashba}, we compute the ambient-pressure band structure by setting the polar mode amplitude to zero, as shown in Fig.~\ref{BS_mode}. Note that even in the absence of the polar mode, the two tilt modes collectively polarise the electron clouds, leading to inversion symmetry breaking and thus preserving the $Bb2_1m$ structure. We find that the absence of the polar mode leads to a substantial enhancement of the Rashba parameter to $\sim$172 meV \AA, which is approximately 1.8 times greater than its value in the relaxed $Bb2_1m$ structure containing the polar mode as well as the two tilt modes (at zero pressure). This suggests that the increasing polar mode amplitude as a function of pressure (discussed in the main text) is responsible for the reduction in the Rashba spin-splitting observed in Fig.~\ref{BS_Rashba}. Additionally, the spin-projected band structures in Fig.~\ref{BS_mode}(b-d) reveal the splitting between spin-up and spin-down bands for the $x$- and $y$- spin-components, confirming the pure Rashba character of the bands at the VBM~\cite{Patel2022}.

\end{document}